\documentclass[a4paper, 10pt, conference]{ieeeconf}      

\IEEEoverridecommandlockouts                              

\usepackage{cite}
\usepackage{amsmath,amssymb,amsfonts}
\usepackage{graphicx}
\usepackage{textcomp}
\usepackage{xcolor}
\usepackage{subfigure}
\usepackage{booktabs}
\usepackage{multirow}
\usepackage[ruled,vlined]{algorithm2e}

\title{\LARGE \bf
Multi-vehicle experiment platform: A Digital Twin Realization Method
}

\author{Chunying Yang$^{1}$ and Jianghong Dong$^{1}$ and Qing Xu$^{1,*}$\\and Mengchi Cai$^{1}$ and Hongmao Qin$^{2}$ and Jianqiang Wang$^{1}$ and Keqiang Li$^{1}$
\thanks{*This work was supported by the National Key Research and Development Program of China under Grant 2019YFB1600804, the National Natural Science Foundation of China under Grant 52072212, Dongfeng Motor Corporation and China Intelligent and Connected Vehicles (Beijing) Research Institute Co.,Ltd. (\emph{Corresponding author: Qing Xu})}
\thanks{$^{1}$ C. Yang, J. Dong, Q. Xu, M. Cai, J. Wang, K. Li are with the School of Vehicle and Mobility, Tsinghua University, Beijing, China 
        {\tt\small ycyacademic@gmail.com,({djh20, cmc18})@mails.tsinghua.edu.cn,({qingxu, sjqlws,likq})@tsinghua.edu.cn}}%
\thanks{$^{2}$H. Qin is with Wuxi Intelligent Control Research Institute, Hunan University, Wuxi, China
        {\tt\small qinhongmao@hnu.edu.cn}}%
}
\begin{document}
\maketitle
\thispagestyle{empty}
\pagestyle{empty}

\begin{abstract}
	With the development of V2X technology, multiple vehicles cooperative control has been widely studied. However, filed testing is rarely conducted due to financial and safety considerations. To solve this problem, this study proposes a digital twin method to carry out multi-vehicle experiments, which uses combination of physical and virtual vehicles to perform coordination tasks. To confirm effectiveness of this method, a prototype system is developed, which consists of sand table testbed, its twin system and cloud. Several aspects are quantified to describe system performance, including time delay and localization accuracy. Finally, a vehicle level experiment in platoon scenario is carried out and experiment results confirm effectiveness of this method.
\end{abstract}

\section{INTRODUCTION}
With the development of V2X communication, coordinated control of multiple vehicles has been widely studied. \cite{ref:lfq} proposed a unified multi-vehicle formation control framework for CAVs (connected and automated vehicles). \cite{ref:ccy} presented a graph-based vehicle cooperation method to guarantee safety and efficiency at unsignalized intersections. Field testing is an effective means of confirming effectiveness of coordination strategies. \cite{ref:filed2} demonstrated that traffic waves can emerge as a result of human behaviors alone via filed experiment . \cite{ref:filed1} conducted small-scale field experiment to prove that stop-and-go waves could be dissipated via control of autonomous vehicles. However, only a limited number of CAVs are used in these filed experiments. 

Digital twin (DT), an emerging representation of cyber-physical systems, has attracted increasing attentions recently. \cite{ref:sum} summarized that DT opened the way to real-time monitoring and synchronization of real-world activities with the virtual counterparts. Compared to traditional methods of simulation and prediction, DT paradigm has the following advantages. First, relying on reliable communication and sensor technologies, virtual models can access real-time data to achieve status update and optimization. Second, since DT paradigm emphasizes two-way interactions between physical and cyber spaces, it is practical for physical elements to receive feedback from its virtual counterparts. \cite{ref:sum} summarized development of DT in industry and analyzed its key enabling technologies and theories. \cite{ref:adas} proposed a modeling architecture for large-scale platform and adopted the architecture for ADAS (advanced driving assistant system) DCU verification. 

In the related research combining DT and CAVs, \cite{ref:wang1} proposed a general DT system framework for connected vehicles, which introduced DT into the field of CAVs for the first time. \cite{ref:wang2} used a DT approach to develop a cooperative ramp merging system and conducted field implementation. Besides, they evaluated system effectiveness in terms of mobility and fuel consumption. However, these researches lack detailed descriptions on how to apply DT paradigm to the field of CAVs.

On the HMI side, \cite{ref:vr} used Web service and augment reality technology to visualize DT. \cite{ref:wang3} and \cite{ref:wang4} built cyber space with game engine and transferred advisory speed that generated in cyber space to driver for tabulate display. 

\cite{ref:idea} proposed a trajectory-following guidance method to chase a virtual target, which inspires method proposed by this study. Relying on real-time information exchanges between physical and cyber spaces, it is practical to carry out multi-vehicle experiment using combination of physical and virtual vehicles.

Compared to other recent study of validation of coordination strategies, contributions of this paper are as follows.
\begin{itemize}
	\item This study proposes a digital twin method to carry out multi-vehicle experiment, which uses combination of physical and virtual vehicles to perform coordination tasks. This method is practical even when available physical vehicles are insufficient. 
	\item A prototype system is developed under digital twin paradigm. Several concerning indices are measured, including time delay and localization accuracy. 
	\item Various devices are provided for the system to enable human-machine interactions (HMI). Mixed reality device (Hololens) is employed for visualization and manipulation of vehicles. Driving simulator allows driver to control vehicles from the first-person perspective.
\end{itemize}

The rest of this paper is organized as follows. Section~\ref{sec:methodology} introduces components and the capabilities of the realization system in detail. Section~\ref{sec:simulation} reveals a case study in platoon scenario and demonstrates system performance in terms of string stability. In the last section conclusion are drawn and future work directions are signaled.
\section{Methodology}

\label{sec:methodology}

In this study, a CAVs-orientated digital twin system is developed in Tsinghua university, which supports carrying out multi-vehicle experiment even when physical vehicles are inadequate. To meet such a purpose, this study proposes using combination of physical and virtual vehicles to carry out experiment. Specifically, a sand table testbed is established, where miniature vehicles can run smoothly. To construct cyber space, game engine is used to complete full-elements modeling. Cyber space can reflect real-time running status of sand table. Moreover, this study introduces the notion of cloud vehicle, which is used to represent miniature vehicles.

The following section describes the system in more technical detail, providing overviews of architecture and components, system performance quantification and HMI designing.
\subsection{System components and capabilities}
\begin{figure}[htbp]
	\includegraphics[width=0.45\textwidth]{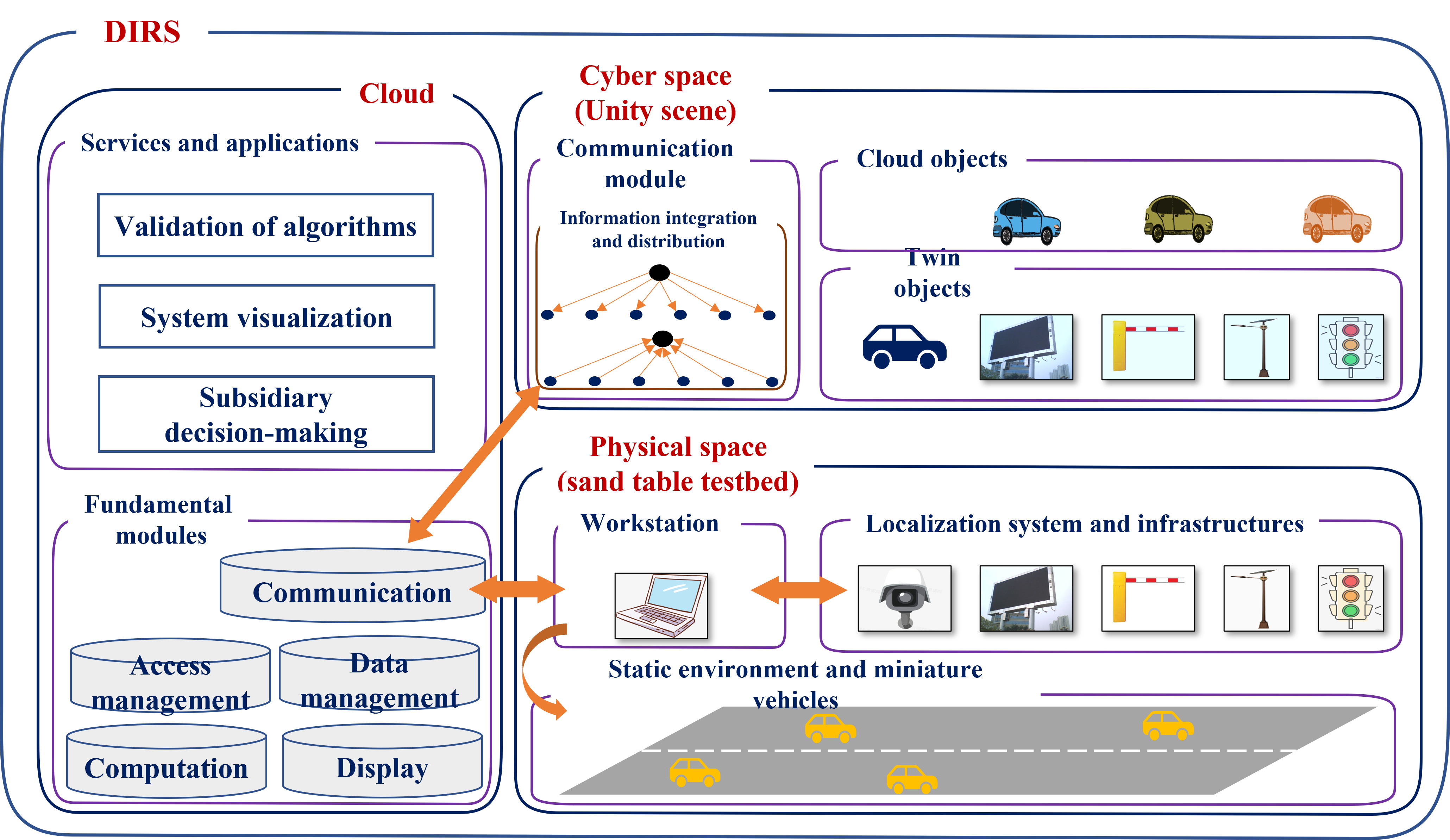}
	\caption{System architecture}
	\label{fig:arc}
\end{figure}
 As depicted in Fig.~\ref{fig:arc}, the system is composed of three major components\,--\,physical space, cyber space and cloud, among which cloud functions as information integration center and links the other two components.
\subsubsection{\textbf{Physical space}}
	Physical space of this system is in the form of sand table testbed as shown in Fig.~\ref{fig:sandTable}, which is provided with structured roads and basic traffic infrastructures. Generally, three miniature vehicles are available for field testing. Each of them is well-configured and can be controlled wirelessly by a nearby workstation. 

\begin{figure}[htbp]
	\begin{centering}
		\subfigure[]{
			\includegraphics[width=0.14\textwidth]{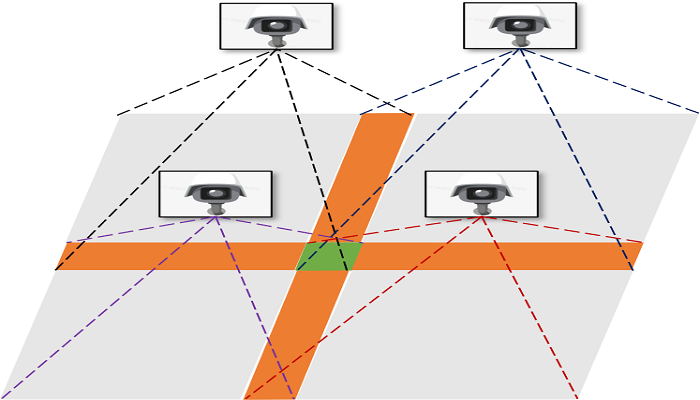}
			\label{fig:cameraCoverage}}
		\subfigure[]{
			\includegraphics[width=0.14\textwidth]{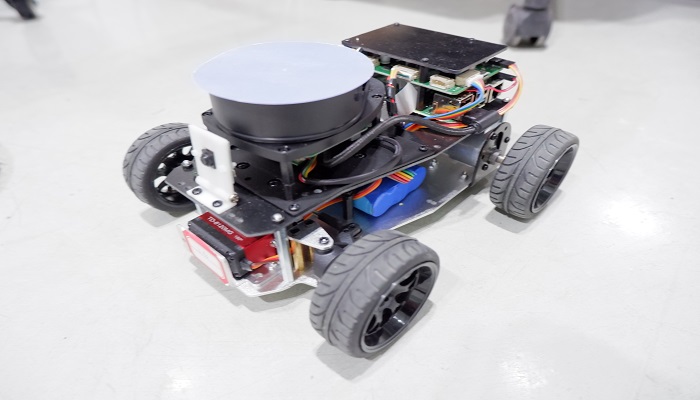}
			\label{fig:vehicle}}
		\subfigure[]{
			\includegraphics[width=0.14\textwidth]{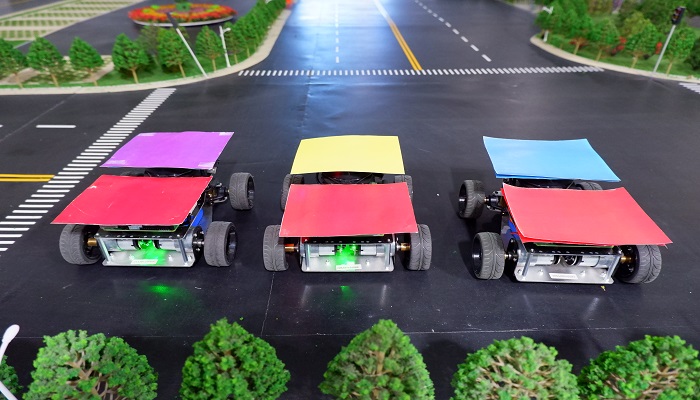}
			\label{fig:coloredVehicle}}
	\end{centering}
	\caption {Physical space. \ref{fig:cameraCoverage} Camera coverage, where white, orange and green parallelograms represent area covered by one, two and four cameras respectively. \ref{fig:vehicle} Miniature vehicle. \ref{fig:coloredVehicle} Vehicles with identifying colored blocks on the top.}
\end{figure}

	Similar to \cite{ref:hotdec}, we choose the miniature vehicle with specific properties in mind. First, steering and speed values can be sent to each vehicle to realize real-time control. Second, control frequency should be no less than $\mathrm{5\,Hz}$, and so improper behaviors or potential damages can be noticed and adjusted in time. Finally, with various sensors and computation unit equipped onboard, the vehicle should possess full-stack capabilities of a CAV, which contain perception, decision-making and actuation. These characteristics lead to the choice of miniature as shown in Fig.~\ref{fig:vehicle}. For individual vehicle, it is equipped with lidar, IMU and on-board camera. 

\begin{figure}[htbp]
	\subfigure[Schematic diagram of vehicle control during trajectory tracking process]{
		\includegraphics[width=0.48\textwidth]{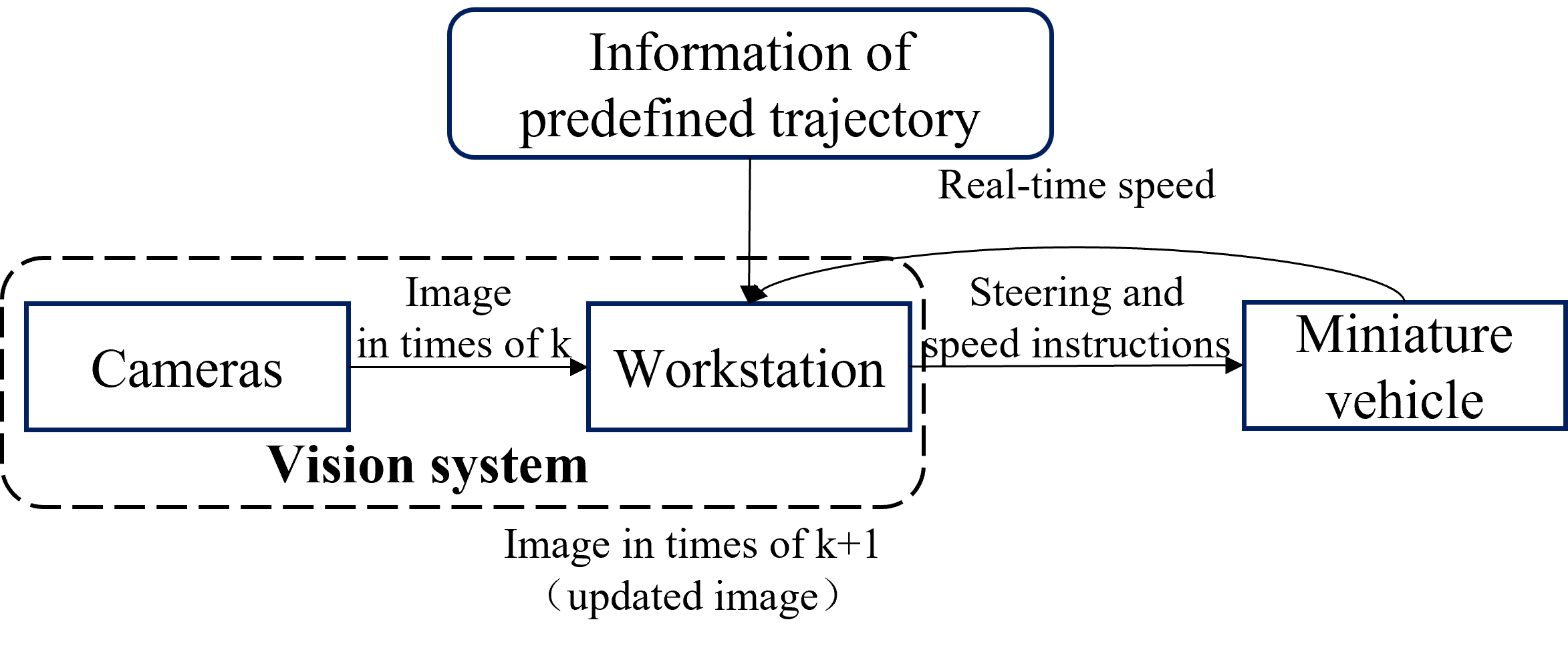}
		\label{fig:vehicleControl}}
	\subfigure[Hardware connections while in operation]{
		\includegraphics[width=0.48\textwidth]{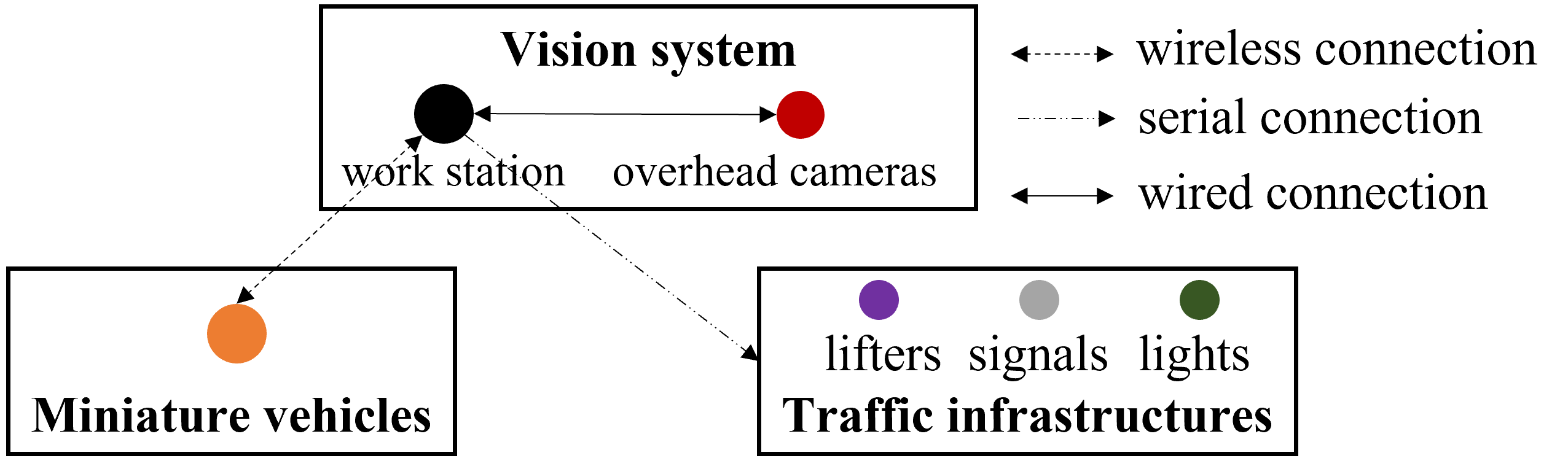}
		\label{fig:connection}}
	\caption{Trajectory tracking process}
\end{figure}

	Trajectory tracking is one of the most common tasks for miniature vehicles. Considering relatively stable light conditions, vision system composed of four overhead cameras is selected to implement localization and orientation measurement. Schematic overview of trajectory tracking process is depicted in Fig.~\ref{fig:vehicleControl} and hardware connections is illustrated in Fig.~\ref{fig:connection}. 

	In order to follow a pre-defined trajectory, workstation sends out steering and speed instructions at frequency of $\mathrm{8\,Hz}$. Once receiving these messages, vehicle will follow these commands to update its status. Overhead cameras capture image of sand table at frequency of $\mathrm{30\,Hz}$, however, output frequency of vision system is around $\mathrm{8\,Hz}$ due to image-processing procedures. Combining information of pre-defined trajectory and real-time data from cameras, workstation will calculate steering and speed instructions of next time for each vehicle to track the desired trajectory.

To realize image-based localization and orientation measurement, several steps are implemented as follows:
\begin{itemize}
	\item Preparation\\
	To identify vehicle, colored blocks with unique combination are attached on the top of each vehicle as demonstrated in Fig.~\ref{fig:coloredVehicle}. Four overhead cameras are reasonably installed to achieve full coverage of the sand table with overlap at the junction. 
	\item Distortion adjustment\\ 
	Distortion is one of inherent characteristics of cameras. Mature algorithms provided by OpenCV is adopted to solve this problem.
	\item Color recognition and pose calculation\\
	Through color space transformation, filtering and extracting, specific colored block can be easily recognized. Since each vehicle has colored blocks at the front and rear, it's quite easy to calculate its position and heading.
\end{itemize}
Due to word limitation, more details about vision system are uploaded to project website\cite{ref:website}.
\subsubsection{\textbf{Cyber Space}}
Several aspects should be taken into account before constructing cyber space of this system.

\begin{figure}[htbp]
	\begin{centering}
		\subfigure[Sand table testbed]{
			\includegraphics[width=0.22\textwidth]{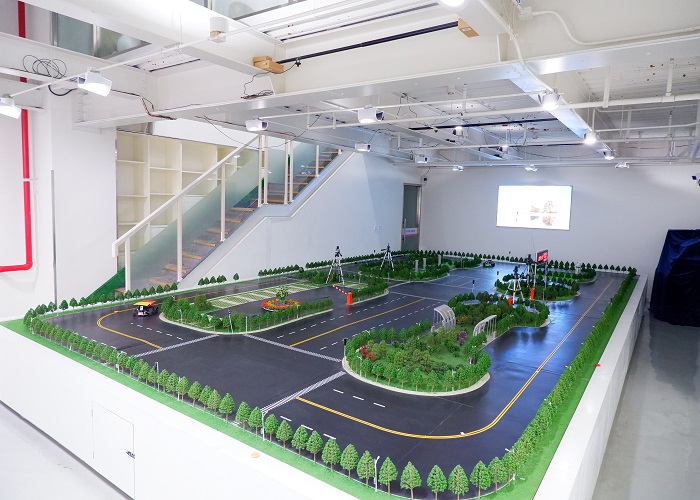}
			\label{fig:sandTable}}
		\subfigure[Cyber space built with Unity]{
			\includegraphics[width=0.22\textwidth]{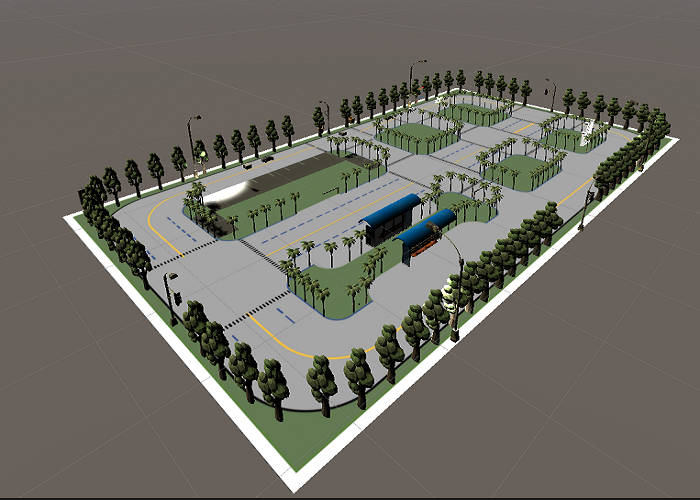}
			\label{fig:cyberSpaceUnity}}
	\end{centering}
	\caption {Physical and cyber counterparts}
\end{figure}

First, cyber space should be capable of reflecting running status of sand table. Second, cyber space is able to provide more immersive experience to support HMI. These traits lead to the choice of game engine Unity3D (Unity) because of its abundant resources. Cyber space built with Unity is shown in Fig.~\ref{fig:cyberSpaceUnity}. 

\begin{figure}[htbp]
	\includegraphics[width=0.43\textwidth]{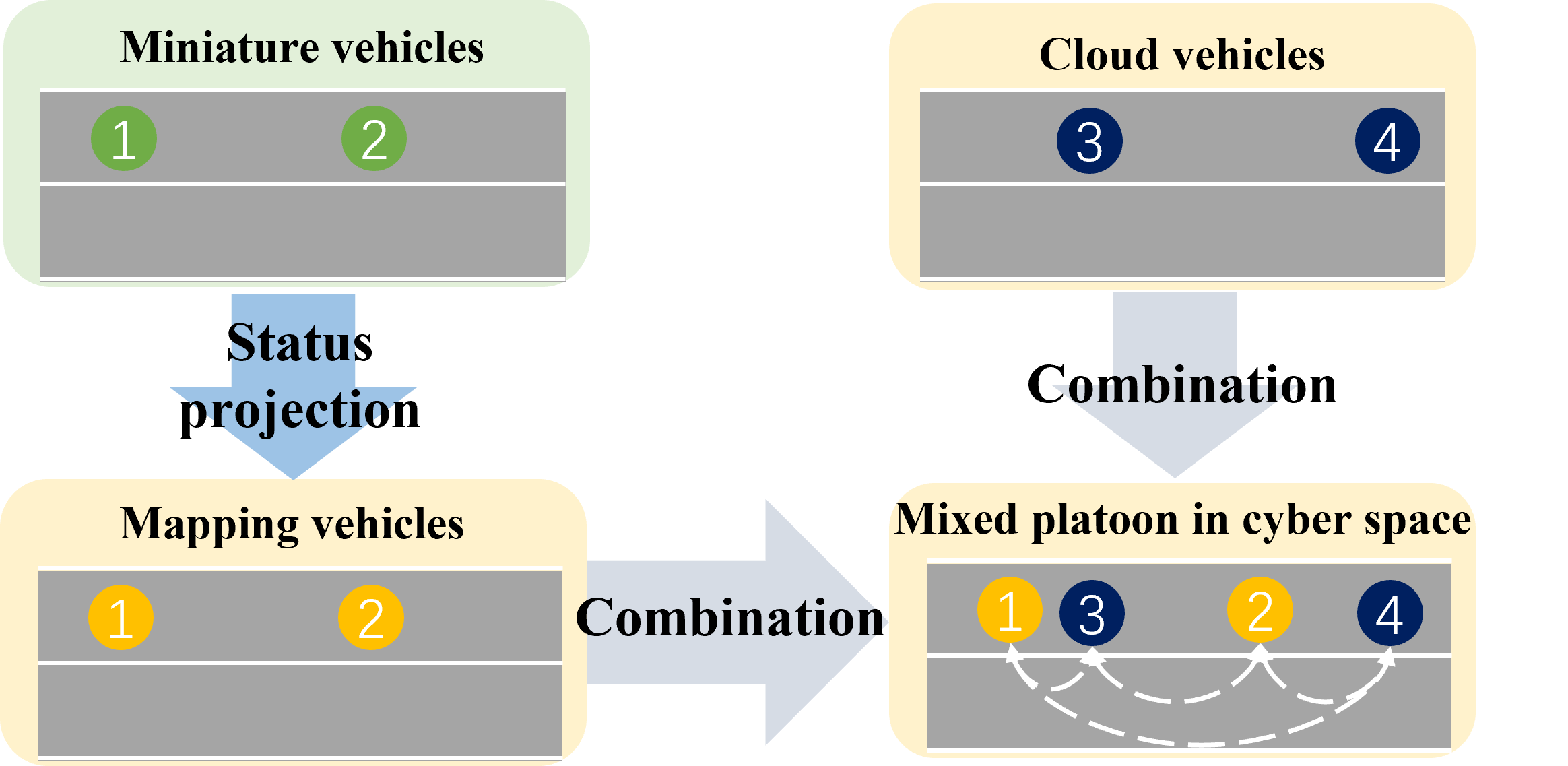}
	\caption{Combination of mapping and cloud vehicles. Status of miniature vehicles can be projected into cyber space through interactions. Since cloud vehicles are exclusively modeled in cyber space, it is technically practical to form the mixed platoon in cyber space.  }
	\label{fig:explain}
\end{figure}

To carry out multi-vehicle experiment without adequate physical vehicles, this study introduces the notion cloud vehicle. Specifically, two types of vehicles are modeled in cyber space---mapping vehicle and cloud vehicle. Mapping vehicle is mainly used to reflect running condition of miniature vehicles. To meet such a purpose, parameters of mapping vehicle, including coordinate, speed and heading, are updated based on real-time information from physical space. Cloud vehicle has no counterpart in physical space and is used to simulate miniature vehicles. Fig.~\ref{fig:explain} explains the idea in detail.

\begin{figure}[htbp]
	\includegraphics[width=0.48\textwidth]{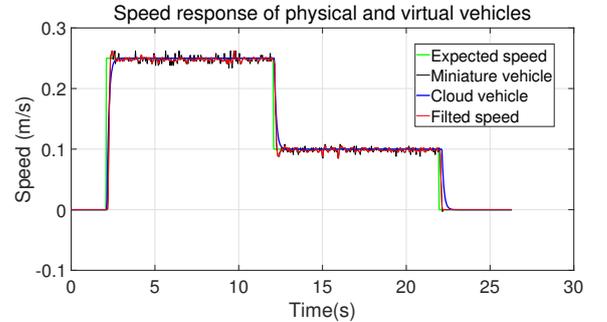}	
	\caption{Speed response of miniature and cloud vehicles.}
	\label{fig:speeds}
\end{figure}

Using cloud vehicle to represent miniature ones is based on the premise that the two have similar dynamic characteristics. Based on multiple experiment results, bicycle model is selected to depict cloud vehicle's dynamic features. Mathematical expression of bicycle model is given by Equation \ref{equ:dynamincs} and corresponding parameters are listed in Table~\ref{tab:equationParas}.
\begin{figure}[htbp]
	\begin{center}
		\includegraphics[scale = 0.5]{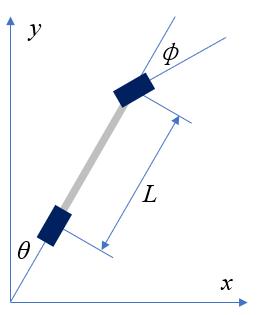}
		\caption{Bicycle model. $L$ represents the wheelbase, $\phi$ represents the steering angle of the front wheel and $\theta$ represents the yaw angle. }
		\label{fig_bicyclemodel}
	\end{center}
\end{figure}
\begin{table}[htbp]
	\centering
	\caption {Vehicle parameters used in this paper.}
	\label {tab:equationParas}
	\begin{tabular}{|c|c|c|c|c|c|}
		\hline
		$L$ & $\phi$ & $\theta$ & $v$ & $a$ \\
		(m) & (degree) & (degree) & (m/s) & (m/$\text{s}^2$) \\
		\hline
		$0.15$ & $[-40,40]$ & $[-180,180)$ & $[0,1]$ & $[-4.5,4.5]$ \\
		\hline
	\end{tabular}
\end{table}
The state of the vehicle contains $X$, $Y$, $\theta$ and $\phi$, where $X$ and $Y$ represent the horizontal and longitudinal position of the center of vehicle's rear axle. The inputs of the vehicle model are $a$ and $\phi$, and the state is calculated as:
\begin{eqnarray}
	\begin{cases}
		\label{equ:dynamincs}
		\dot{X}=v\cdot sin(\theta)\\
		\dot{Y}=v\cdot cos(\theta)\\
		\dot{\theta}=tan(\phi)\cdot v/L\\
		\dot{v}=a\\
	\end{cases}
	\label{equ:model}
\end{eqnarray}
Given the same expected speed at the same time, Fig.~\ref{fig:speeds} shows speed responses of miniature and cloud vehicles.

\subsubsection{\textbf{Cloud}}
\begin{figure}[htbp]
	\includegraphics[width=0.45\textwidth]{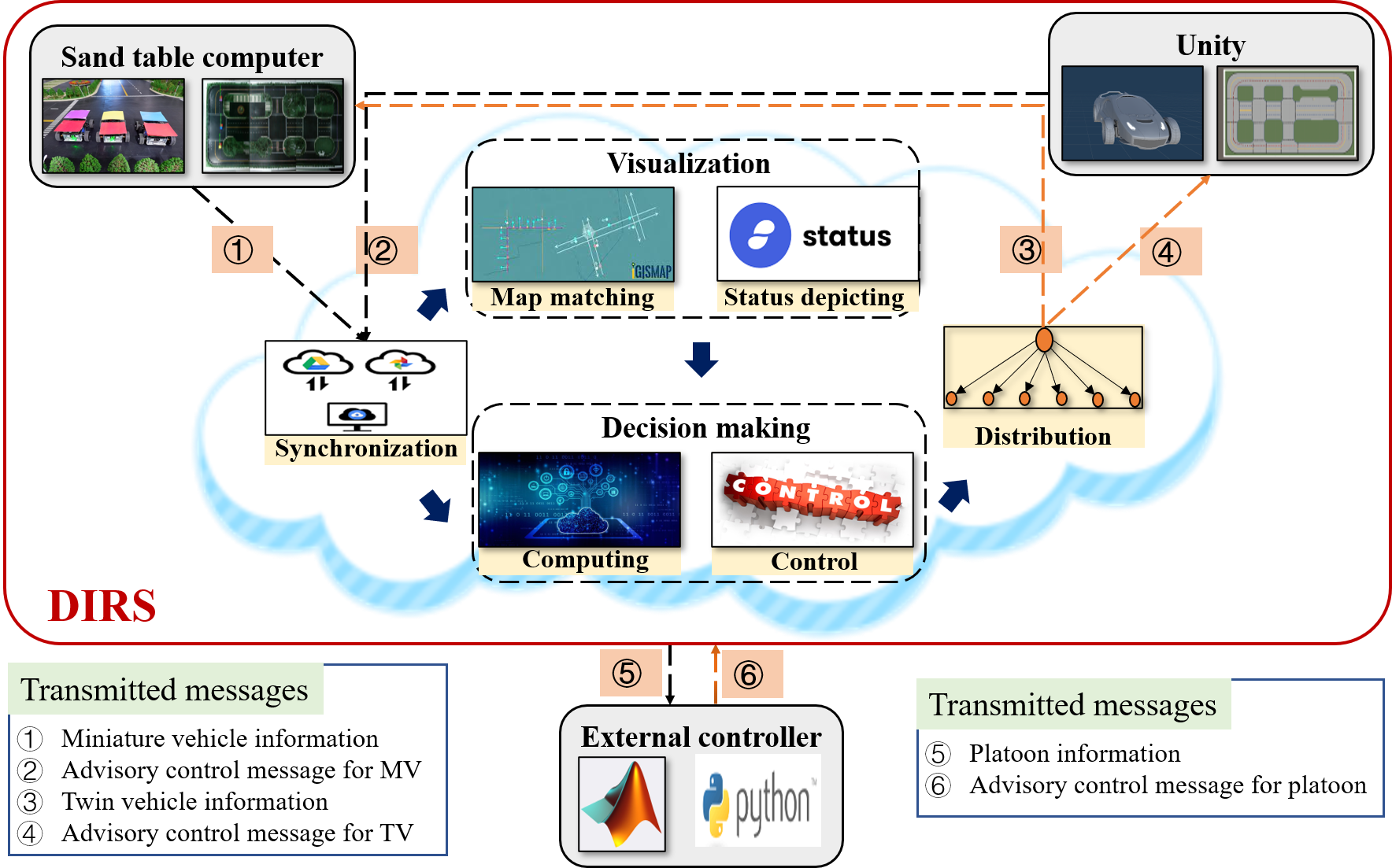}
	\caption{Functions of cloud}
	\label{fig:cloud}
\end{figure}

One of major goals in developing the system is to provide a platform where various algorithms or methods can be evaluated without paying attention to internal implementation process. Designing cloud is motivated by high demands of access and computation. As shown in Fig.~\ref{fig:cloud}, cloud collects data from both physical and cyber spaces, it is feasible for external controller or program to access system data by establishing connection with cloud. Also, cloud provides various services to external program, for example, system visualization.

\begin{figure}[htbp]
	\includegraphics[width=0.48\textwidth]{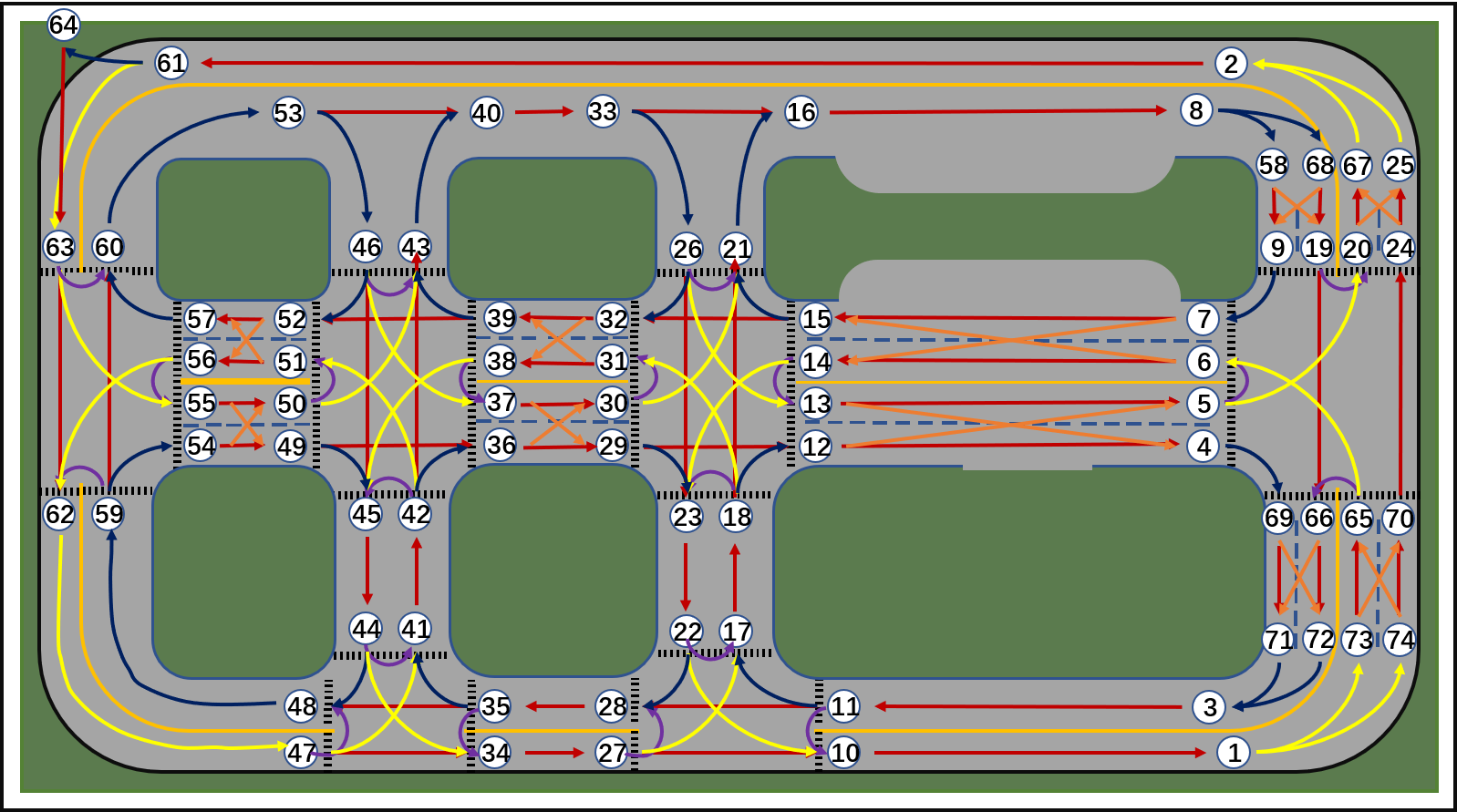}
	\caption{Collection of calibrated nodes}
	\label{fig:calibratedNodes}
\end{figure}

When an external controller or program is connected to cloud for vehicle controlling, three modes are provided and each of them opens up different control parameters as follows:
\begin{itemize}
	\item  Speed and steering can be assigned to each vehicle directly. Once receiving these values, the sandbox mechanism of miniature vehicles will be launched to guarantee that these values will not exceed its valid range. Besides, sandbox mechanism makes decision on control priorities. 
	\item Sequence of equidistant way points. In this study, way point refers to discrete point of a desired trajectory. Sequence of way points can be assigned to each vehicle to follow the trajectory, where vehicle calculates values of speed and steering autonomously.
	\item Node, which denotes the divergence or merging point of vehicle platoon, can be assigned to each vehicle. Fig.~\ref{fig:calibratedNodes} presents all nodes calibrated with unique coordinate.
\end{itemize}

Modes above are designed to facilitate accesses with different purposes. For instance, study of CAVs-orientated formation control can assign nodes for each vehicle as target lane or position, while study that requires vehicle to follow a desired trajectory can assign sequence of way points to it. 

\subsection{System performance quantification}
As a open system to facilitate experiment, several aspects should be taken into account, such as time delay and localization accuracy. Several parameters and quantification results are listed as follow to present more technical details of the system. Important static parameters are listed in Table~\ref{tab:summarize}.

\begin{table}[htbp]
	\centering
	\caption {Static parameters}
	\label {tab:summarize}
	\begin{tabular}{|c|c|}
		\hline
		\textbf{Parameters}&\textbf{Value} \\
		\hline
		Size of sand table&9\,m$\times$5\,m\\
		Width of lane&240mm\\
		Mass of vehicle&1.4\,kg\\
		Size of chassis&200mm$\times$180mm$\times$130mm\\
		Resolution of overhead camera&1920$\times$1080\\
		Resolution of onboard camera&640$\times$480\\
		Frame rate of overhead camera&30fps\\
		Maximum speed of vehicle&1\,m/s\\
		Maximum steering angle of vehicle & 40°\\
		Maximum control frequency of vehicle& 20Hz\\
		\hline
	\end{tabular}
\end{table}

System dynamic parameters listed in this section include: time delay, localization and mapping accuracies. Since these factors affect system performance directly, it is necessary to quantify them precisely. 
\subsubsection{Delay of time}
\begin{figure}[htbp]
	\includegraphics[width=0.48\textwidth]{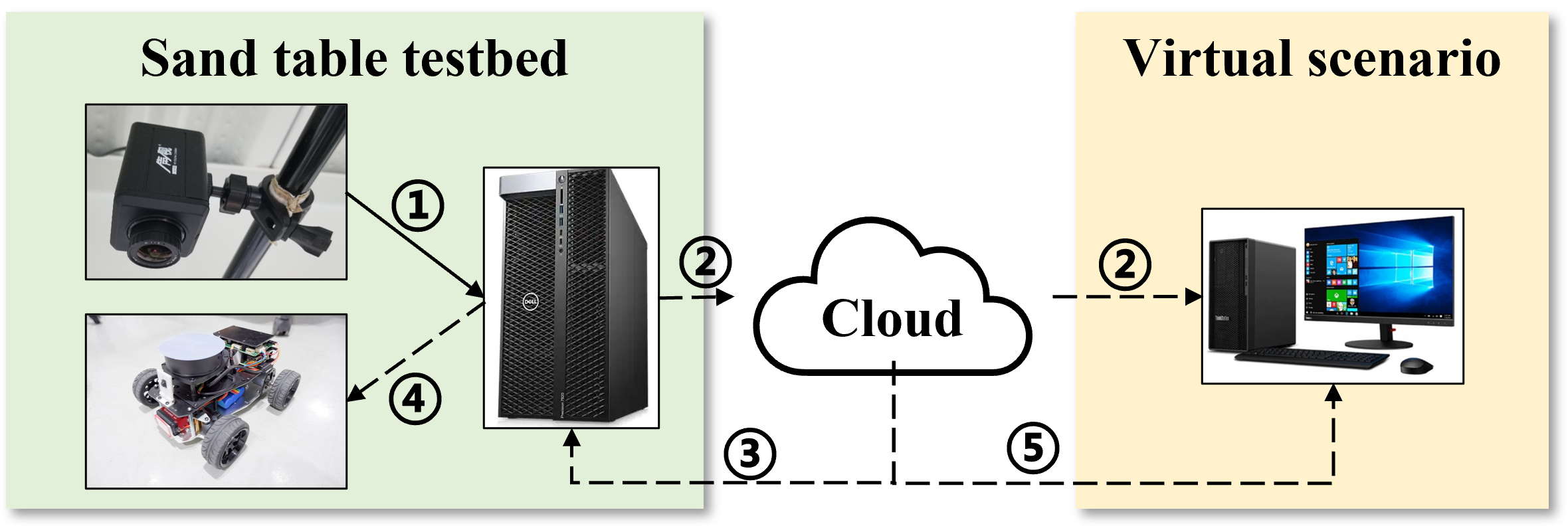}
	\caption {Main stages of time delay in the system. Stage$\,1$ denotes data flow from camera to a nearby workstation and stage$\,2$ denotes data flow from workstation to the device that runs Unity, where cloud functions as an immediate agent. Stage 4 denotes data flow from workstation to miniature vehicles. Stage 3 describes data flow from cloud to workstation and stage 5 describes data flow from cloud to device runs Unity.}
	\label{fig:delay}
\end{figure}

\begin{figure}[htbp]
	\begin{centering}
		\subfigure[Stage1]{
			\includegraphics[scale=0.08]{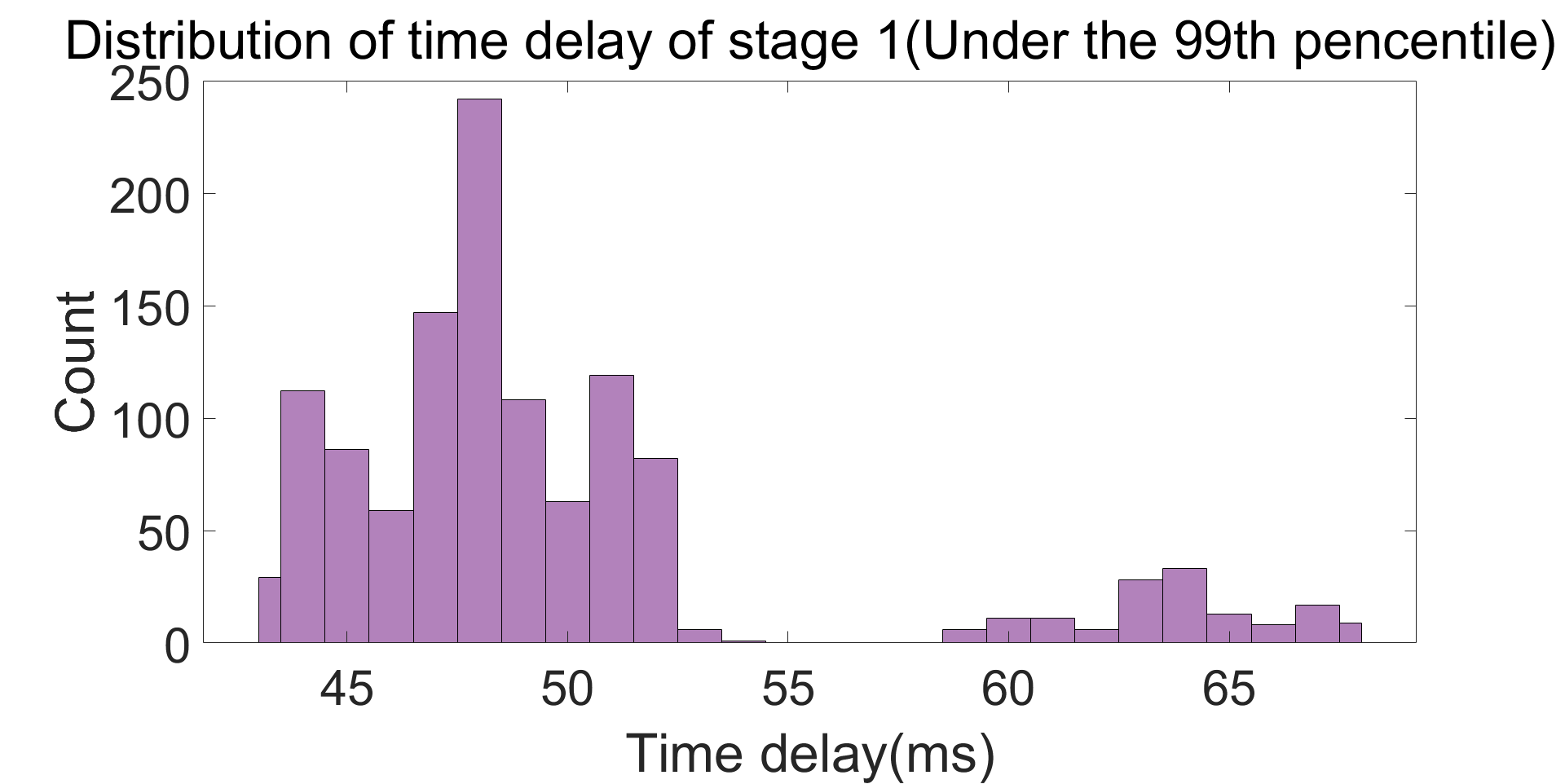}}
		\subfigure[Stage2]{
			\includegraphics[scale=0.08]{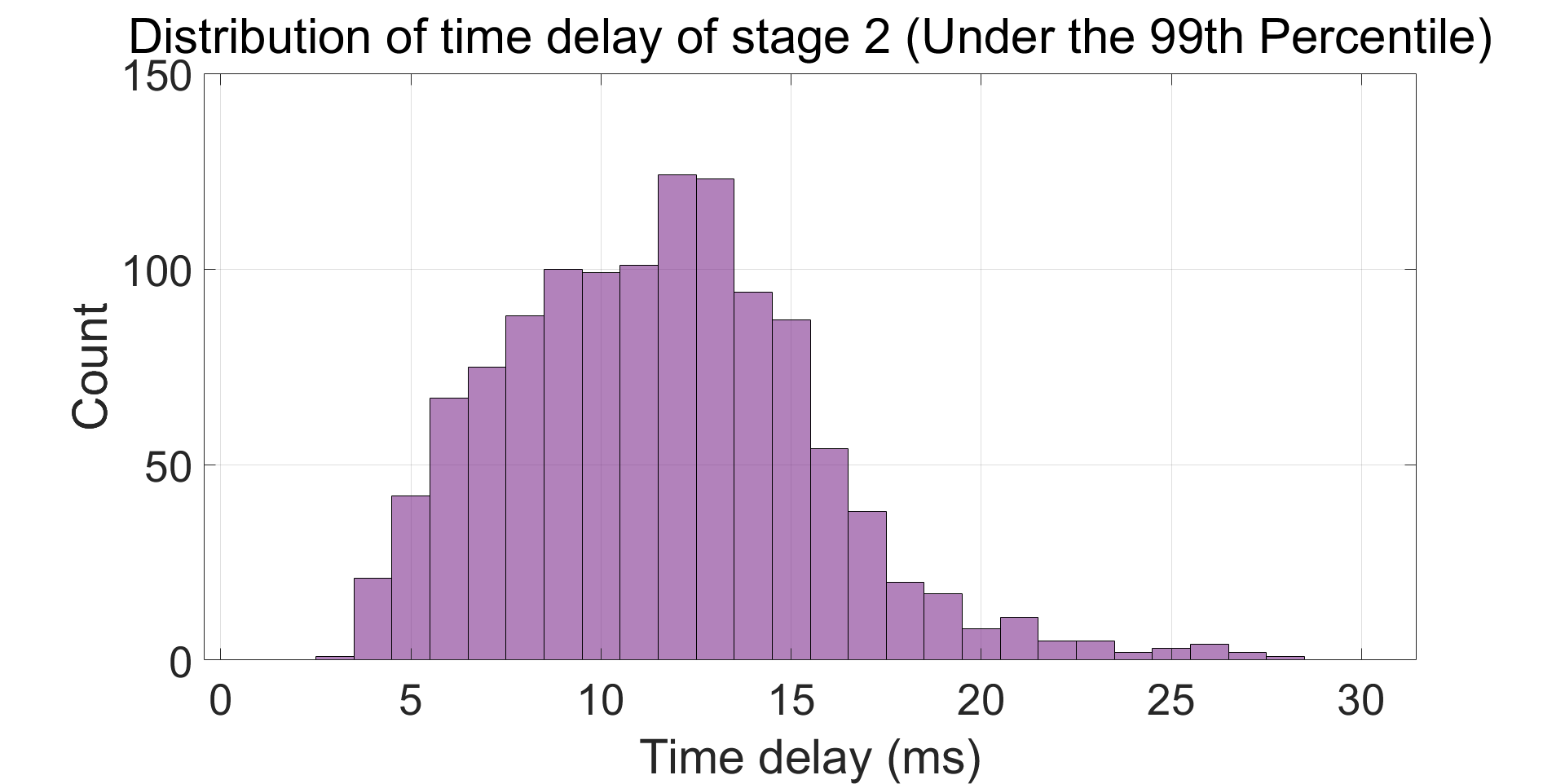}}
		\subfigure[Stage3 and stage5]{
			\includegraphics[scale=0.08]{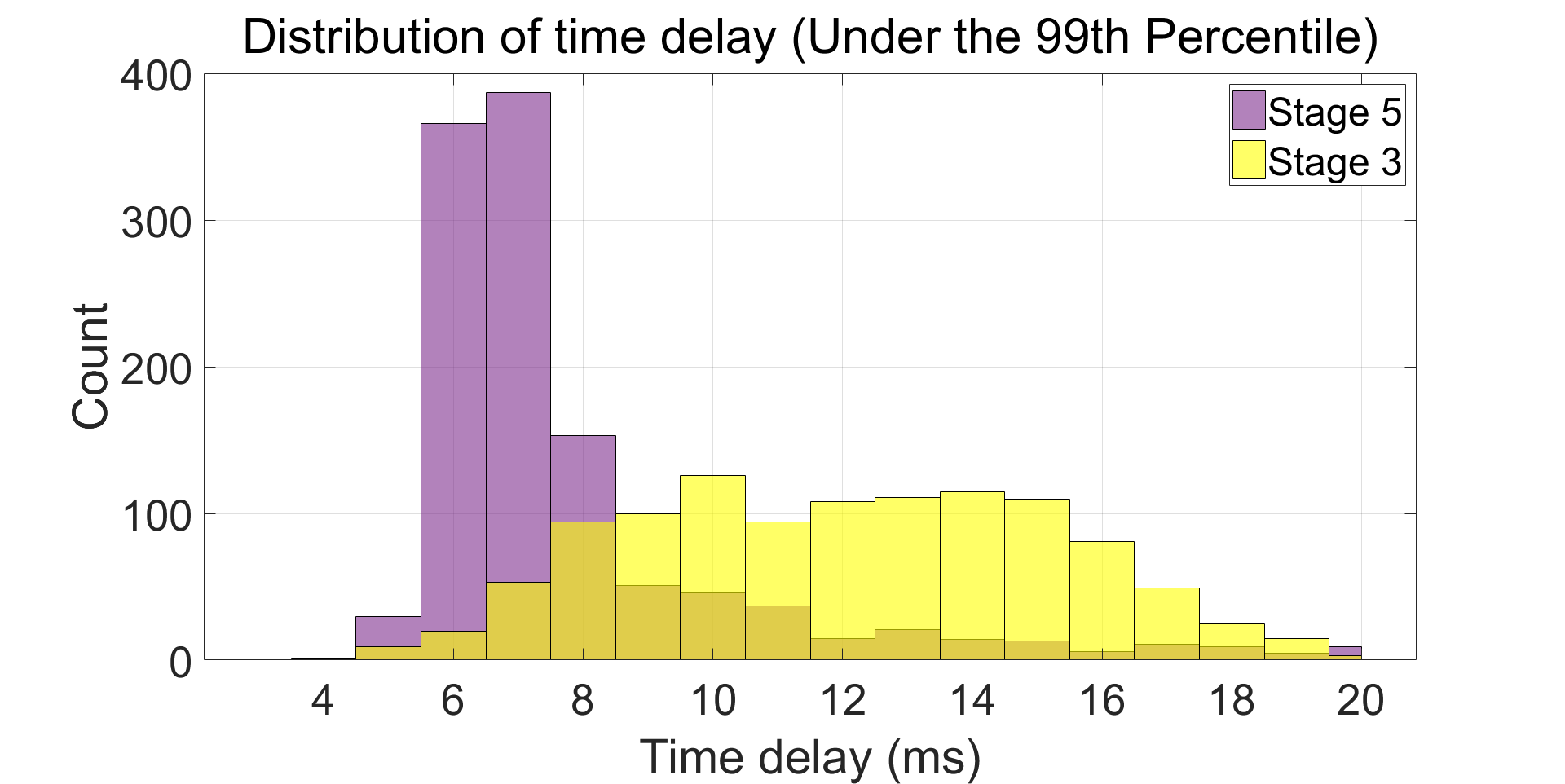}}
		\subfigure[Stage4]{
			\includegraphics[scale=0.08]{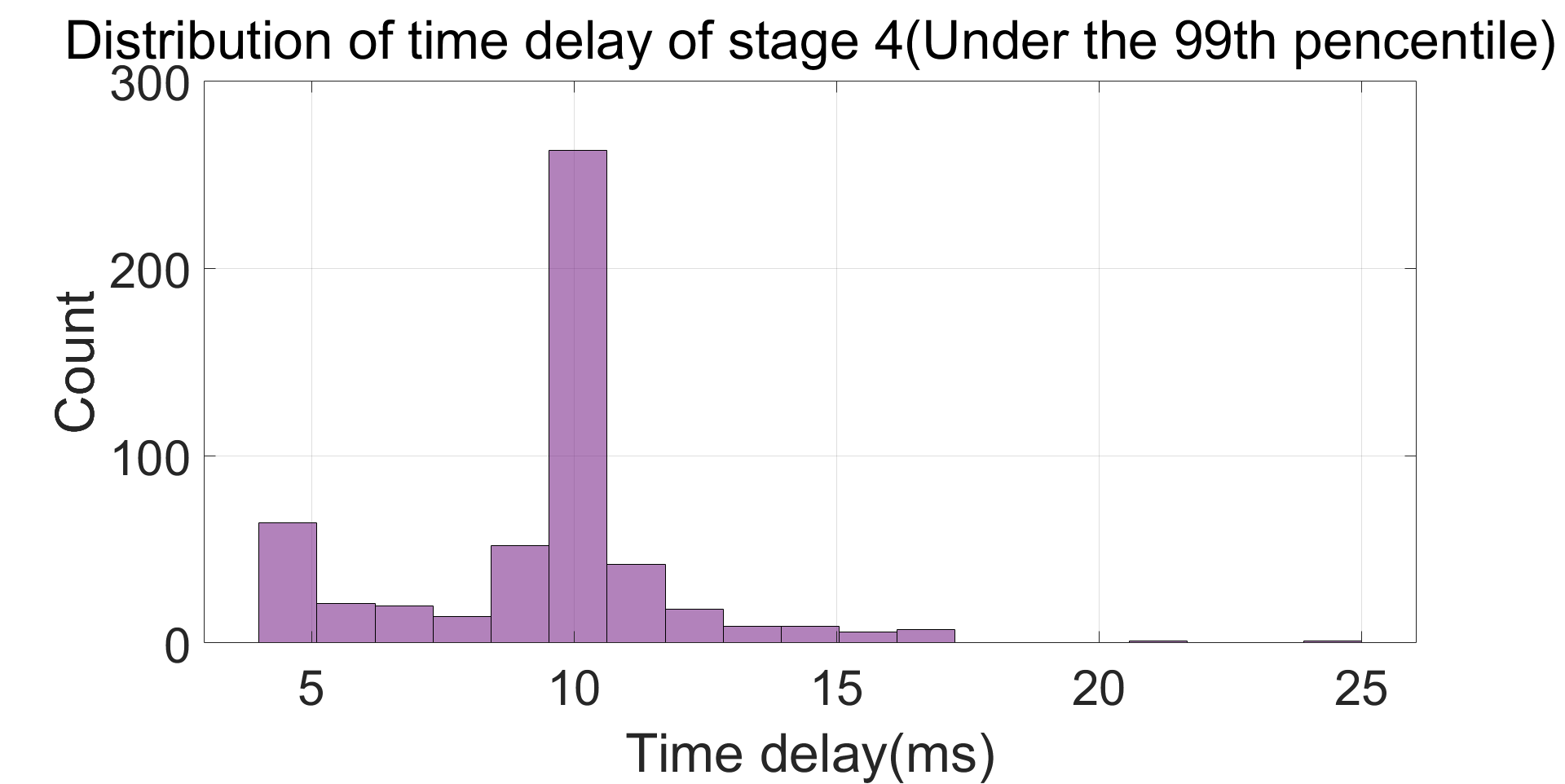}}
	\end{centering}
	\caption {Distribution of time delay for each stage. }
	\label{fig:distribution}
\end{figure}
Time delay is a critical hardware property of the system. As depicted in Fig.~\ref{fig:delay}, there are five main stages of time delay in the system. Time delay of stage$\,1$ is defined by time difference of image transmission process. Stage$\,1$ is caused by electronics in cameras. Camera takes pictures at frequency of $30$\,Hz and is set to transmit the picture before capturing another. Since process above is repetitive, time delay of stage$\,1$ can be measured by recording duration between frame intervals. 
 
Time delay of stage$\,2$ is defined by time duration of message transmission process. Assuming that workstation of physical space sends message to cloud at times of $t_1$. When cloud receives message at times of $t_2$, it will combine information from other nodes before sending outwards. Device running Unity receives processed message at times of $t_3$. Network lantency of stage$\,2$ is calculated as $t_3-t_1$. Compared to delay of stage$\,1$, this delay is effected by many other factors like number of access nodes and network quality. As a result, it is hard to quantify it accurately because cloud and the two devices may not locate within the same network and results vary a lot when it happens. In the simplest case that cloud and the two devices locate within the same network and are connected wirelessly, an experiment is conducted to quantify it. Similarly, time delay of stage$\,3$ and $5$ is highly effected by the same factors. Experiment is conducted when these devices locate within the same network.

Delay of stage$\,4$ is a crucial index of this system, since it is related to control performance directly. Time delay of stage$\,4$ is defined as duration of message transmission. Experiment is conducted when workstation and vehicles are in the same network. 

Fig.~\ref{fig:distribution} represents distribution of these five stages and Table~\ref{tab:statistics} summarizes statistic data.
\begin{table}[htbp]
	\centering
	\caption {Statistical results of the five stages}
	\label {tab:statistics}
	\begin{tabular}{ccccccc}
		\hline
		Stage&Sample size& Average & Max& Min& 99$^{th}$ pencentile\\
		&&$(ms)$&$(ms)$&$(ms)$&$(ms)$\\
		\hline
		$1$&1200&49.7&69&39&67 \\
		$2$&1200&11.7&61&3&26\\
		$3$&1200&14.7&97&4&66\\
		$4$&641&8.4&61.4&2.9&16\\
		$5$&1200&8.3&141&4&23\\
		\hline
	\end{tabular}
\end{table}

\subsubsection{Localization and mapping accuracy}
Since position is one of the inputs of vehicle's control strategy in many cases, it is necessary to quantify localization accuracy. Since mapping errors result from localization error theoretically, and therefore mapping errors should be in accordance with localization errors.

To quantify localization accuracy, errors along X axis (parrallel to short side of sand table) and Y axis (parrallel to short side of sand table) are measured within operational areas. Each area is a square of length $0.5\,m$. Average localization error along X axis is 17.1$\,mm$ while maximum value is 41$\,mm$. Average localization error along Y axis is 14.9$\,mm$ while maximum value is 43$\,mm$. 

\subsection{System display and HMI designing}
\label{subsec:hmi}
\begin{figure}[htbp]
		\subfigure[View from Hololens]{
		\label{fig:hololens}
		\includegraphics[width=0.22\textwidth]{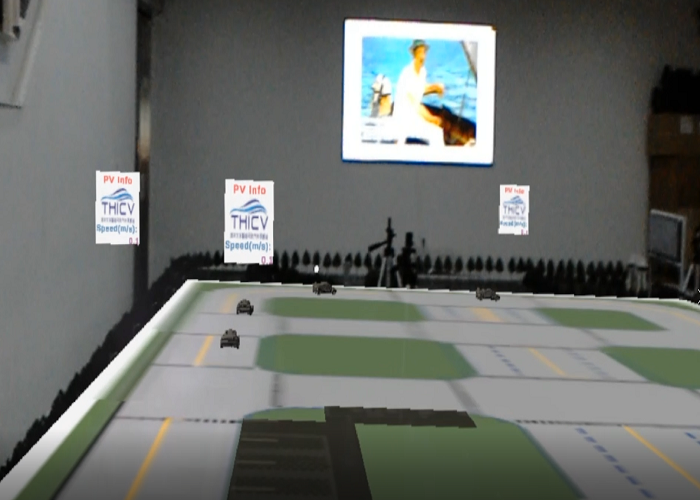}}
		\quad
		\subfigure[People wearing Hololens]{
		\includegraphics[width=0.22\textwidth]{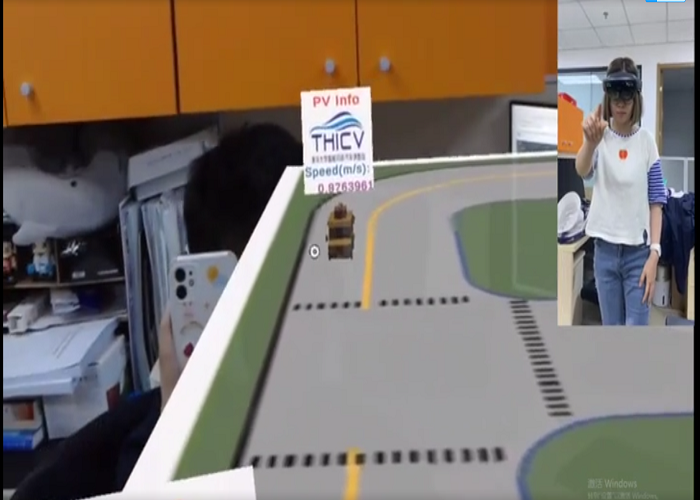}}
		\quad
		\subfigure[Driving simulator]{
		\includegraphics[width=0.22\textwidth]{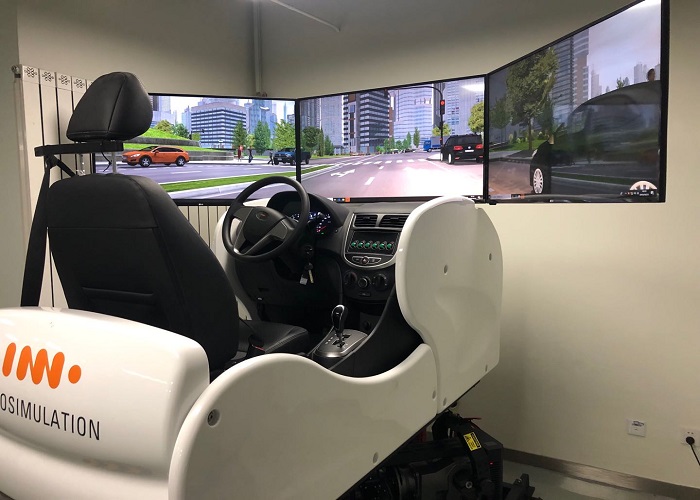}}
		\quad
		\subfigure[People manipulating simulator]{
		\includegraphics[width=0.22\textwidth]{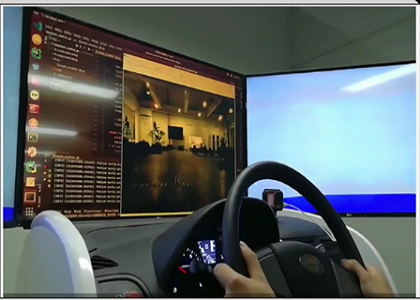}}
	\caption {Several HMI methods designed for DIRS}
	\label{fig:hmi}
\end{figure}

Many studies has been carried out to visualize cyber space and to enable HMI. In this study, we desire to visualize cyber space in a novel way and to get people more involved. As a result, several devices are provided to support various interaction methods: mixed reality device (Hololens), driving simulator and screen display, which can be seen in Fig. \ref{fig:hmi}.

Scenes rendered by Hololens are in the form of realistic holograms. People wearing Hololens can manipulate objects of cyber space easily, for example, speeding up a virtual vehicle. Through communication and information exchanges, it is technically practical for Hololens to manipulate miniature vehicles as well. Hololens is leveraged mainly to render overlap of the sand table and its virtual counterpart as shown in Fig.~\ref{fig:hololens}. Through proper configuration, cloud vehicle and miniature vehicle can run together on surface of sand table. Driving simulator provides first-person perspective of one of these vehicles and applies people's control input onto it. Screen display provides one of the most intuitional presentations of cyber space and also provides convenient interface to adjust observation perspective.
\section{Case study and result evaluation}
\label{sec:simulation}

To confirm system effectiveness, one vehicle level experiment is carried out in this section and corresponding results are presented.

\subsection{ Scenario Description}

\begin{figure}[htbp]
	\includegraphics[width=0.48\textwidth]{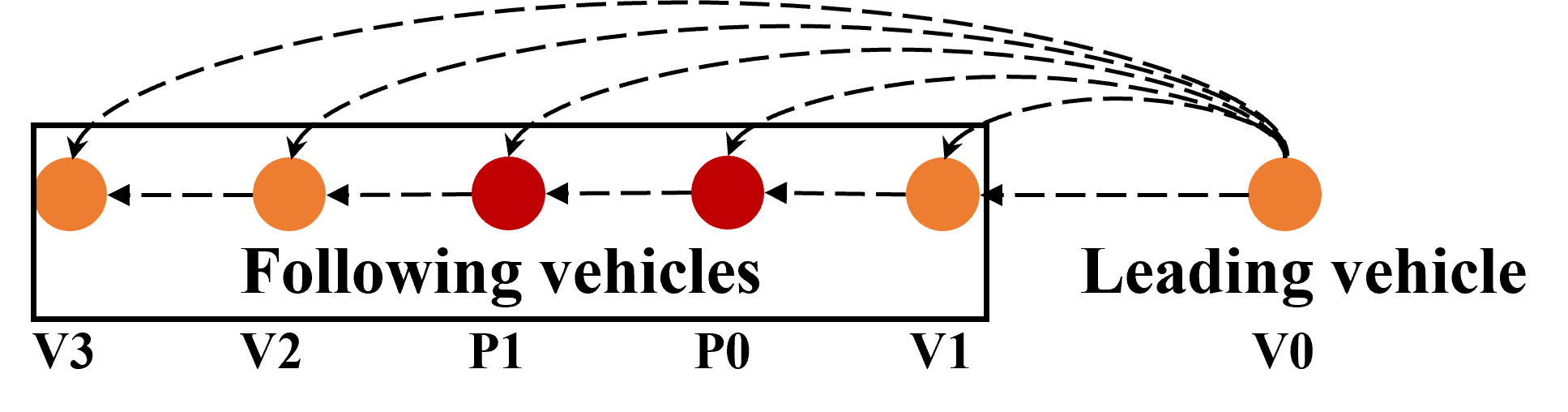}
	\caption{Initial formation of platoon, which contains all following relationships between vehicles---V follows V, V follows P, P follows V and P follows P, where P denotes physical miniature vehicle and V denotes virtual cloud vehicle.}
	\label{fig:topology}
\end{figure}

Scenario that platoon drives in a rounded rectangular loop is designed to form the mixed platoon and to analyze different interaction types between vehicles. To evaluate system performance, scenario configuration is set as follows. 
\begin{itemize}
	\item The outer ring road of sand table is selected as desired trajectory for the platoon. 
	\item Initial formation of the platoon is shown in Fig.~\ref{fig:topology}, which considers all following relationships between different types of vehicles. 
	\item Speed of leading vehicle shows periodic changes. To guarantee safety, maximum speed of miniature vehicle and cloud vehicle are set to $0.26\,m/s$ and $0.3\,m/s$ respectively.
	\item During implementation process, preview control is adopted as lateral control strategy. Longitudinal control considers both spacing maintaining (\textbf{0.5m}) and speed keeping, which is given as:
	\begin{eqnarray*}
		\label{equ:pidcontrol}
		\begin{cases}
			a_i=k_{pi}(X_i-X_{\text{i-1}})+k_{v1}^i(v_i-v_{\text{l}})+k_{v2}^i(v_i-v_{\text{i-1}})\\
			v_i^{k+1}=v_i^k+a_i*\Delta$t$\\
		\end{cases}
	\end{eqnarray*}
	where $acc_i$ is the control input of vehicle $i$ and $k_{pi}$, $k_{v1}^i$ and $k_{v2}^i$ are feedback gains. $X_{\text{i-1}}$ and $v_{\text{i-1}}$ are longitudinal spacing and speed of its preceding vehicle. $v_l$ is speed of leading vehicle.
\end{itemize}

\subsection{Result analysis}
\begin{figure}[htbp]
	\begin{centering}
		\subfigure[Speed profiles]{
			\label{fig:platoon_speeds}
			\includegraphics[width=0.48\textwidth]{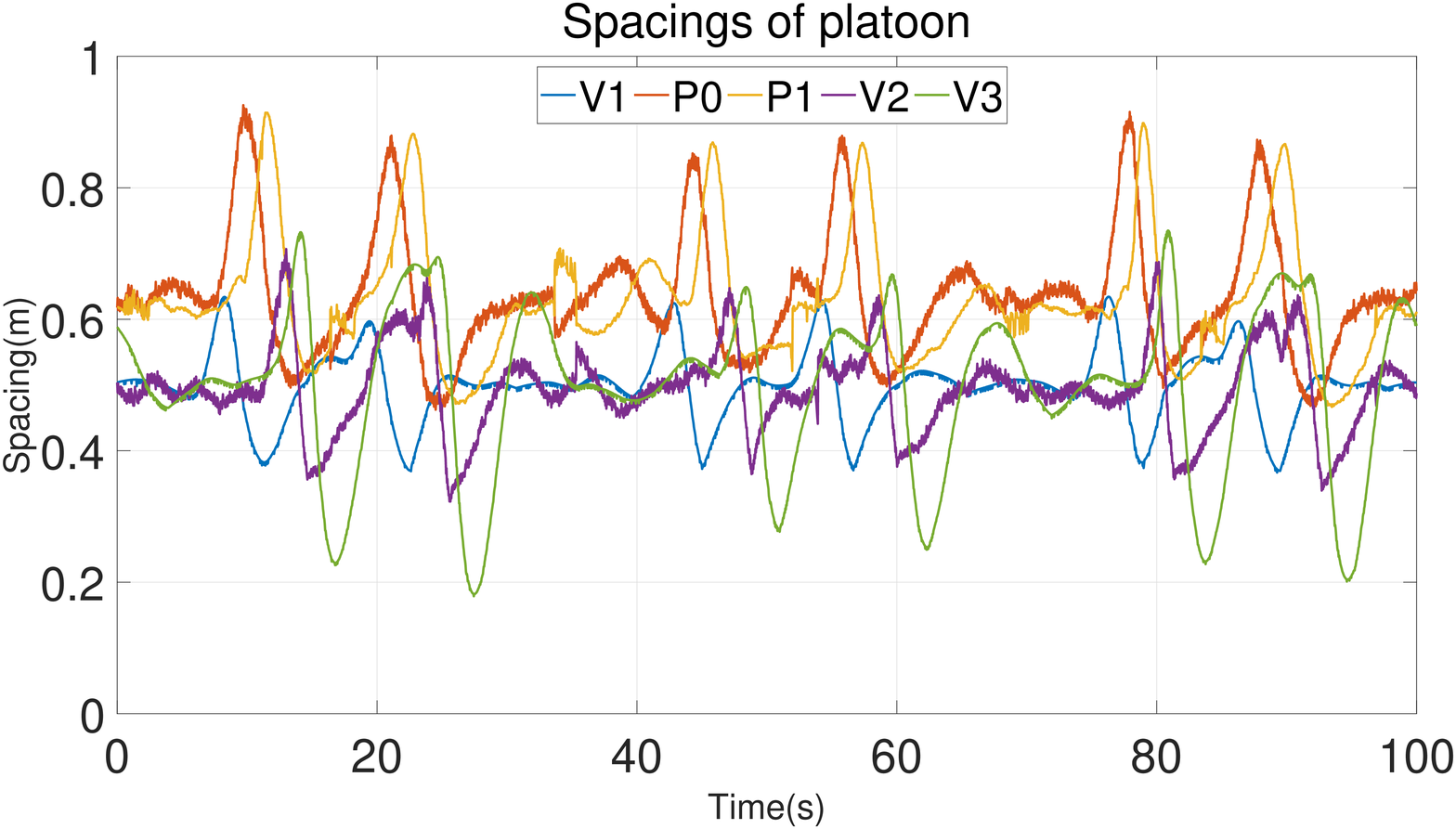}}
		\subfigure[Spacing profiles]{
			\label{fig:platoon_spacing}
			\includegraphics[width=0.48\textwidth]{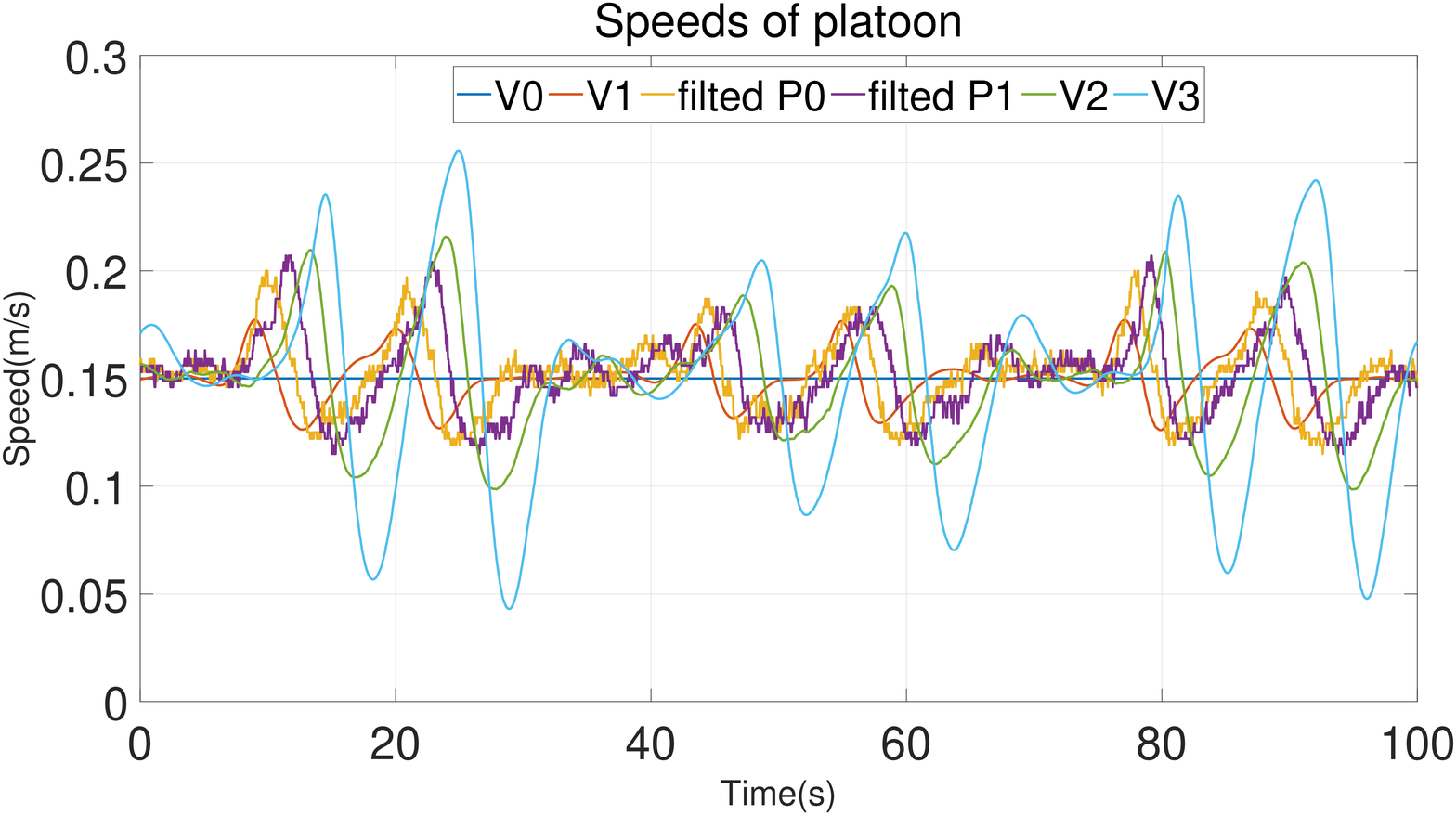}}
	\end{centering}
	\caption{Experiment results}
	\label{res:platoon}	
\end{figure}
\vspace{-3mm}
Fig.~\ref{fig:platoon_speeds} displays speed profiles of each member, based on which several conclusions can be drawn. First, compared to physical vehicles (in red and yellow), speed changes of virtual ones are much more smoother. Second, speed profiles between the same type of vehicles show a relatively high degree of similarity while speed profiles between different types of vehicle vary a lot. Finally, platoon remain relatively stable without any stop-go-waves. Fig.~\ref{fig:platoon_spacing} describes spacing profiles of each following member to its proceeding vehicle, based on which several characteristics are summarized. First, spacing changes of each following member remain relatively stable (mostly varies between 0.9m and 1.3m). Second, to follow a physical vehicle, spacing variations of V2 (cloud vehicle, in purple) remain unsmooth, which is similar to that of miniature ones (in orange and blue). 

\section{Conclusion and future work}
\label{sec:conclusion}
To carry out multi-vehicle experiment when available real vehicle are insufficient, this study proposes a digital twin method to solve this problem. Moreover, a prototype system is developed. The prototype system is mainly composed of sand table testbed, its twin system and cloud. To enrich interactivity of the system, several novel ways are designed to support HMI, including mixed reality device (Hololens), driving simulator and screen display. An experiment in platoon scenario is conducted and results are analyzed, which confirms that this system is capable of providing an open platform to carry out multi-vehicle experiment. Further more, several aspects need more considerations along its future application: 1) Impact of different access modes on experiment results will be analyzed. 2) More dynamics models of cloud vehicle will be provided.

\end{document}